\documentclass{statsoc}
\usepackage[a4paper]{geometry}
\usepackage[utf8]{inputenc}
\usepackage{amsmath}
\usepackage{amssymb}
\usepackage[hidelinks]{hyperref}
\usepackage{xr}
\usepackage{ulem}
\usepackage{textcomp}
\usepackage{etoolbox}
\usepackage{tikz}
\usetikzlibrary{arrows, automata}

\makeatletter
\newcommand*{\addFileDependency}[1]{
  \typeout{(#1)}
  \@addtofilelist{#1}
  \IfFileExists{#1}{}{\typeout{No file #1.}}
}
\makeatother

\newcommand*{\myexternaldocument}[1]{%
    \externaldocument{#1}%
    \addFileDependency{#1.tex}%
    \addFileDependency{#1.aux}%
}

\myexternaldocument{web_materials}

\makeatletter
\patchcmd{\@makecaption}
  {\parbox}
  {\advance\@tempdima-\fontdimen2} 
  {}{}
\makeatother 

\title[Causal feature selection for spatiotemporal disease risk mapping]{Nonparametric Causal Feature Selection for Spatiotemporal Risk Mapping of Malaria Incidence in Madagascar}
\author[]{Rohan Arambepola}
\address{Oxford Big Data Institute, 
Nuffield Department of Medicine, 
University of Oxford}
\email{rohan.arambepola@stx.ox.ac.uk}
\author[]{Peter Gething}
\address{Oxford Big Data Institute, 
Nuffield Department of Medicine, 
University of Oxford;
\\Telethon Kids Institute, Perth Children's Hospital, Perth, Australia;
\\and Curtin University, Perth, Australia}
\author[R. Arambepola, P. Gething and E. Cameron]{Ewan Cameron}
\address{Oxford Big Data Institute, 
Nuffield Department of Medicine, 
University of Oxford;
\\Telethon Kids Institute, Perth Children's Hospital, Perth, Australia;
\\and Curtin University, Perth, Australia}

\usepackage[caption=false]{subfig}
\usepackage{graphicx}
\usepackage{ifthen}
\usepackage{bm}
\usepackage{float}
\usepackage{booktabs}
\usepackage{xcolor}
\usepackage[export]{adjustbox}
\usepackage{placeins}

\newcommand{\indep}[3]{
    \ifthenelse{\equal {#3} {}}{#1\perp\!\!\!\perp#2}{#1\perp\!\!\!\perp#2 | #3}
}

\begin{document}

\maketitle
\begin{abstract}
 Modern disease mapping draws upon a wealth of high resolution spatial data products reflecting environmental and/or socioeconomic factors as covariates, or `features', within a geostatistical framework to improve predictions of disease risk.  Feature selection is an important step in building these models, helping to reduce overfitting and computational complexity, and to improve model interpretability.  Selecting only features that have a causal relationship with the response variable could potentially improve predictions and generalisability, but identifying these causal features from non-interventional, spatiotemporal data is a challenging problem.  Here we examine the performance of a causal feature selection procedure with regard to estimating malaria incidence in Madagascar.  The studied procedure designed for this task combines the PC algorithm with spatiotemporal prewhitening and kernel-based independence tests extended to accommodate aggregated data.  This case study reveals a clear advantage for causal feature selection in terms of the out-of-sample predictive accuracy in a forward temporal estimation task, but not in a spatiotemporal interpolation task, in comparison with thresholded spike-and-slab, for both linear and non-linear regression models. Compared to no feature selection, causal feature selection was most beneficial in settings wherein the volume of available data was low relative to the model complexity.
\end{abstract}

\section{Introduction}
Spatial mapping of malaria risk is an important public health tool, facilitating the efficient allocation of limited resources and the precision targeting of interventions~\citep{elliot2000spatial, lawson1999disease, drake2017geographic}. While maps produced to-date have typically focussed on annual summaries of risk (e.g.\ \citealt{bhatt2015effect}), there is increasing interest in modelling at higher temporal resolutions~\citep{colborn2018spatio,  haddawy2018spatiotemporal, nguyen2019statistical}. Monthly or weekly risk maps have the potential to improve the timing of control strategies that rely on knowledge of seasonal trends to interrupt transmission, such as indoor residual spraying and seasonal chemoprevention~\citep{griffin2016potential, landier2018spatiotemporal}, and to aid in real-time identification of outbreaks for malaria early warning systems~\citep{girond2017analysing, minakawa2018establishment, pan2018challenges, tompkins2018dynamical}.

Modern disease mapping methods combine population health metrics from routine surveillance systems and/or cross-sectional surveys with ancillary spatial data products to produce high resolution risk maps within a probabilistic framework~\citep{bhatt2013global, bhatt2015effect, gething2016mapping, kang2018spatio}.  These spatial data products may be based on satellite imaging, weather station data, or third-party model-based outputs and may reflect multiple aspects of the local geography relevant to disease transmission.  Due to the wealth of such products available as potential covariates some form of feature selection is important to avoid overfitting and to improve predictive performance.  Feature selection is particularly challenging when mapping at higher temporal resolutions, as the possibility of including each product at a number of time lags can result in a large number of often highly correlated features to choose from. For example, when mapping annual incidence each dynamic covariate (such as rainfall) corresponds to a single potential feature (an annual summary of this covariate), while when predicting monthly incidence in this analysis each dynamic covariate generates four features (each at a different time lag). Substantial variations observed in the seasonal pattern of malaria transmission, even in nearby locations~\citep{singh2000seasonality} or year-to-year in the same locations~\citep{reiner2015seasonality}, suggest a complex system of interactions between environmental variables, socioeconomic factors, and disease risk.  Hence, a general feature selection algorithm for malaria risk mapping must be applicable to both linear and non-linear regression models.

Classical feature selection procedures aim to maximise model fit or predictive performance with respect to the observed data---as measured, for example, through cross-validation-based metrics or information criteria---while penalising the size of the feature set; typically retaining features with large associations with the response variable or those that are strong predictors within the model framework~\citep{guyon2003introduction, liu2007computational}.  In contrast, causal feature selection procedures~\citep{guyon2007causal} aim to select features that have a direct causal relationship to the response variable even if the association is comparatively weak, excluding features that are only associated with the response variable due to common causes or through mediating variables.  Causal feature selection has been applied to several real-world datasets across diverse application areas including photo-voltaic cell engineering, financial time series forecasting, and medical prognosis, with mixed results: improvements in predictive accuracy over non-causal feature selection approaches in some cases~\citep{hmamouche2017causality, sun2015using, zhang2014causal}, but not others~\citep{cawley2008causal}. 

Many causal inference algorithms depend on conditional and unconditional independence tests~\citep{spirtes2000causation}, but the most commonly-used independence tests require strong assumptions (such as joint Normality) which may not hold true when using spatiotemporal environmental and epidemiological data. Furthermore, routine malaria incidence data, like many other sources of epidemiological data, are often aggregations of cases over geographical areas. For example, case counts from health facilities can be thought of as an aggregation of cases occurring in the catchment area of that facility (the area from which the patients are drawn). This is a problem as relationships between variables on an aggregate level may not hold on a fine-scale; a scenario known as the `ecological fallacy' \citep{wakefield2006health}. Here we apply spatiotemporal prewhitening~\citep{flaxman2016gaussian} and independence tests based on the theory of kernel embeddings~\citep{muandet2017kernel} within the PC algorithm for graph discovery~\citep{spirtes1995learning, spirtes2000causation} to perform nonparametric causal feature selection in an explicit spatiotemporal setting. We utilise the kernel embeddings framework to extend kernel-based independence tests to infer fine-scale relationships from the aggregated incidence data, in a similar way to distribution regression methods~\citep{flaxman2015supported}. These features are used in a Bayesian hierarchical model for estimating malaria incidence in Madagascar, and the results compared to those from the same model fit without feature selection and with a thresholded spike-and-slab model (a common non-causal method;~\citealt{mitchell1988bayesian, george1993variable, kuo1998variable, ishwaran2005spike}).

\subsection{Malaria risk mapping}
Visualisation of raw epidemiological and/or entomological data (e.g. health facility case numbers or prevalence of malaria from health surveys) is an important first step in characterising spatiotemporal patterns in malaria risk for a given area, but these data may suffer biases and incompleteness due to low treatment-seeking rates and incomplete record keeping (amongst other factors).  Relying on raw data is also problematic in pre-elimination settings, where the low transmission rates lead to a very noisy sampling distribution for standard malaria metrics.  Modern disease mapping uses a formal statistical framework combining random field models with high resolution covariates to achieve spatial smoothing and interpolation of raw data~\citep{ribeiro2001geor, diggle2013spatial, bhatt2013global, shearer2016estimating, gething2011modelling, bhatt2015effect}. These maps are used for allocation of resources, both on a global and local scale, and for precision targeting of interventions~\citep{world2019world}.

\subsection{Malaria in Madagascar}
Malaria is a major public health problem in Madagascar with an estimated 2.16 million cases occurring in 2018~\citep{world2019world}.  After a large reduction between 2000 and 2010, reported case numbers have been steadily increasing.  Although perhaps partly due to improvements in access to rapid diagnostic capabilities, these data are nevertheless believed to reflect in some part a genuine increase of malaria transmission in the country~\citep{howes2016contemporary}. This interpretation is supported by spatiotemporal models of community-based infection prevalence data from cross-sectional health surveys~\citep{kang2018spatio}.  In the medium and high transmission zones of Madagascar the clinical incidence rate typically follows a seasonal pattern peaking in April-May, although an earlier February peak has been observed in some places.  Annual trends are less consistent in low transmission areas where instead temporal variation displays an outbreak dynamic~\citep{howes2016contemporary, randrianasolo2010sentinel}, with unusual climatic events and changes in intervention coverage previously identified as key risk factors~\citep{kesteman2016multiple}.  The dominant species of malaria parasite on the island is \textit{Plasmodium falciparum}, although \textit{Plasmodium vivax} is also endemic in many areas~\citep{howes2018risk, kesteman2014nationwide}.

\subsection{Causal feature selection}
It is well known that an association between two variables $X$ and $Y$ in observational data does not necessarily imply a causal effect (``correlation does not equal causation''), due to the possible existence of confounding variables (that is, other variables which have a causal effect on both $X$ and $Y$). Part or all of the observed relationship between $X$ and $Y$ could arise from these common causes.  The ``gold standard'' for discovering the causal effect of a variable $X$ on variable $Y$ is through a randomised control trial (RCT), as the randomisation removes any outside causes of $X$ and thereby removes all confounding pathways. However, there are many situations in which RCTs are not possible (for example, when investigating the causal effect of climatic variables) and instead only `opportunistic', or `retrospective', observational data is available. Causal inference methods (e.g.\ \citealt{rubin05,pearl09}) aim to discover causal relationships and effects in this setting.

While understanding causality is not always necessary for effective prediction, selecting features based on causal relationships might be desirable for a number of reasons. Causal feature selection could produce much smaller feature sets in situations where the response variable is correlated with many features but directly caused by relatively few~\citep{guyon2007causal}.  Smaller feature sets may reduce overfitting and allow for the use of more flexible models.  Causal feature selection could also improve the interpretability of predictive models, where it is often hoped that structure of the fitted model is a reflection of the fundamental mechanistic properties of the system~\citep{guyon2007causal}.  Models built on causal feature sets may also be more robust to problems encountered in predictive modelling, such as concept drift and covariate shift~\citep{scholkopf2012causal}, ideally improving accuracy in estimates made forward in time or in previously unobserved locations. Finally, nonparametric causal selection can be thought of as a form of model-free feature selection and therefore may be used in combination with `over-parameterised', machine learning-style models (the non-linear regression setting) for which classical feature selection methods (such as the LASSO; \citealt{tibshirani1996regression}) may not be applicable.

\subsection{Causal discovery algorithms}\label{intro:causal_discovery}
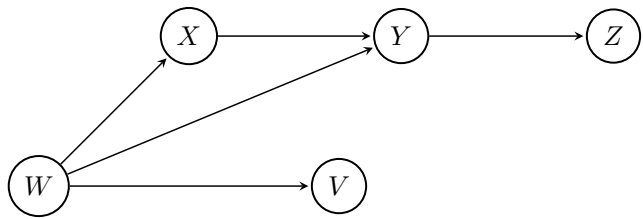
\begin{figure}
\begin{center}
\begin{tikzpicture}[
            > = stealth, 
            shorten > = 1pt, 
            auto,
            node distance = 3cm, 
            semithick 
        ]
        \tikzstyle{every state}=[
            draw = black,
            thick,
            fill = white,
            minimum size = 4mm
        ]
        \node[state] (X) {$X$};
        \node[state] (Y) [right of=X] {$Y$};
        \node[state] (Z) [right of=Y] {$Z$};
        \node[state] (W) [below left of=X] {$W$};
        \node[state] (V) [below right of=X] {$V$};
        \path[->] (X) edge node {} (Y);
        \path[->] (Y) edge node {} (Z);
        \path[->] (W) edge node {} (X);
        \path[->] (W) edge node {} (Y);
        \path[->] (W) edge node {} (V);
        
\end{tikzpicture}
\end{center}
\caption{An example of a DAG.}
\label{fig:DAG_example}
\end{figure}

Causal discovery algorithms aim to infer causal relationships between random variables from observational data, rather than data from designed experiments or RCTs~\citep{glymour2019review, pearl09}. These algorithms assume that the true causal structure between the variables can be represented as a directed acyclic graph (DAG) in which nodes represent variables and edges represent direct causal relationships. For example, the DAG in Figure \ref{fig:DAG_example} represents a causal structure where $W$ is a direct cause of $X$, $V$; $W, X$ are direct causes of $Y$; and $Y$ is direct cause of $Z$. 

There are two main classes of causal discovery algorithms, score-based and  constraint-based methods. Score-based methods, such as the Greedy Equivalence Search~\citep{chickering2002optimal}, associate candidate DAGs with models and use some measure of model fit (such as the Bayesian Information Criterion) to evaluate the DAG. An important advantage of these methods is that they admit a Bayesian approach which quantifies uncertainty in the causal discovery process. However, these methods typically need strong assumptions on the distributions of observed variables and the relationships between them  \citep{heckerman1999bayesian}. A notable exception is the method proposed by~\cite{huang2018generalized} which, like our work, makes use of kernel-based independence tests.

Here we focus on constraint-based algorithms. These algorithms use two assumptions, the Causal Markov Condition and Faithfulness \citep{spirtes2000causation}, to generate a set of conditional independence relationships corresponding to any DAG. The Causal Markov Condition states that a variable is independent of any of its non-descendants given its parents. Here a variable $A$ is said to be a parent of variable $B$ if there exists an edge $A \rightarrow B$, and a variable $D$ is a descendent of variable $C$ if there exists a directed path $C\rightarrow...\rightarrow D$. For example, in Figure \ref{fig:DAG_example}, $Z$ is conditionally independent of $X, W$ or $V$ given $Y$. Similarly, $Y$ is conditionally independent of $V$ given $X, W$. This fundamental condition connects causal graphs with the probability distributions they generate, allowing a statistical approach to causal discovery. The second assumption, Faithfulness, is that the only (conditional or unconditional) independence relationships in the joint distribution are those implied by the Causal Markov Property. Constraint-based algorithms use independence relationships observed in the data to reduce the number of possible underlying DAGs, rejecting DAGs which are not consistent with these relationships. For example, the DAG in Figure \ref{fig:DAG_example} is not consistent with data in which $Y$ and $Z$ appear to be unconditionally independent (due to Faithfulness) or where $X$ and $Z$ are dependent conditional on $Y$ (by the Causal Markov Condition). Constraint-based algorithms are advantageous because they are based entirely on the implications of these two fundamental graphical assumptions. In practice, common independence tests (such as testing for vanishing partial correlations) impose further distributional assumptions on the relationships between variables. However, as we will see, this can be avoided by using non-parametric tests.

\section{Methods}
\subsection{Data}
Monthly malaria case data between January 2013 and December 2016 were available from the National Malaria Control Program of Madagascar (NMCP). These data represent individuals who seek care at a health facility where a diagnosis of malaria is made and where that case is then recorded and entered into the national surveillance system. Such data are termed `passive' case detection and rarely capture all malaria incidence in the community, as some malarious individuals might not seek care or may do so only from informal or private providers. These data do not differentiate by the species of malaria parasite causing the infection. Data were available from 3342 health facilities in Madagascar, of which 2801 were geolocated (NMCP, pers comm) and used for the present study. The facilities that could not be geolocated are believed to be smaller basic health centres and these 541 facilities accounted for only 5.4\% of total cases observed at all facilities over the four year period. Therefore the vast majority of the reported cases were included in our analysis. While the precise location of these facilities was not known, the district in which they were located was. The distribution of these facilities (shown in supplementary material Figure S1) was fairly even, with a higher concentration in the capital Antananarivo. 

The spatial data products gathered to assist with probabilistic interpolation and forecasting of the incidence rates reflected in the routine case data are summarised in Table \ref{cov_table}. These are a mixture of static variables (fixed, non-temporal; e.g. elevation) and dynamic (temporal; e.g. land surface temperature) variables. Dynamic variables were included at 0, 1, 2, and 3 month time lags to allow covariate selection to identify the optimal lag (or lags) to include in the final model. All of these variables are commonly used in malaria risk mapping and have putative causal connections to malaria incidence, however the relative importance of these factors in any given location and (for dynamic variables) the exact time scales on which they act is not necessarily known (see, e.g., \citealt{weiss2015re}). In total there were 40 features available---8 static and 32 dynamic (each of the 8 dynamic variables at 4 different time lags). In the rest of this paper, we continue to use `feature' to mean either one of the static variables or one of the dynamic variables at a specific time lag (for example, we refer to rainfall as a variable and rainfall at a 1 month time lag as a feature). The potential time lags considered were based on the nature of the disease, as symptom onset often occurs between 7 days and 4 weeks of infection. This time period may be similar to the true time lag if the variable considered affects vector biting behaviour (such as temperature). If the variable affects the availability of vector breeding sites (such as rainfall) then the time lag of the effect could be several weeks more.

\renewcommand{\arraystretch}{1.2}
\begin{table}
\caption{\label{cov_table}List of covariates}
\centering
\begin{tabular}{p{13em}p{15em}p{4em}} \toprule
Covariate & Description  & Type\\ \midrule
Accessibility~\citep{weiss2018global} & Distance to the nearest city with population~$>$50,000 & Static\\
Aridity~\citep{trabucco2009global} & Aridity index; ratio of rainfall to evapotranspiration rate, calculated from historical (1970-2000) climatic data & Static\\
Elevation~\citep{farr2007shuttle} & Elevation as measured by the Shuttle Radar Topography Mission & Static\\
PET~\citep{trabucco2009global} &Potential Evapotranspiration; estimated capacity of the landscape to move water from the surface and soil to the air, calculated from historical (1970-2000) climatic data & Static \\
Slope~\citep{farr2007shuttle} & Average slope of the pixel, determined from the elevation of pixels in its neighbourhood & Static\\
Night Lights~\citep{elvidge2017viirs} & Index that measures the ambient luminosity of towns, cities and other sites with persistent lighting & Static\\
DistToWater~\citep{lehner2004global, dtw2} & GIS-derived surface that measures distance to permanent and semi-permanent water based on presence of lakes, wetlands, rivers and streams, accounting for slope and precipitation& Static \\
TWI~\citep{farr2007shuttle} &Topographic wetness index; slope-based estimate of local drainage capacity & Static \\
CHIRPS~\citep{funk2014quasi} & Rainfall & Dynamic\\
LST day~\citep{modisLST} & Daytime land surface temperature & Dynamic \\
LST night~\citep{modisLST} & Night-time land surface temperature & Dynamic \\
TCB~\citep{modisTCB} & Tasselled cap brightness; spectral measure of land reflectance, associated with ground cover type & Dynamic\\
EVI~\citep{modisEVI} & Enhanced vegetation index; spectral estimate of vegetation coverage and `lushness' & Dynamic\\
TSI Pf~\citep{weiss2014air} & Temperature suitability index for \textit{P. falciparum} & Dynamic\\
TSI Pv~\citep{gething2011modelling} & Temperature suitability index for \textit{P. vivax} & Dynamic
\end{tabular}
\end{table}

\subsection{Modelling incidence}\label{subsec:incidence_model}
In this section, we describe the geostatistical model used to interpolate and forecast incidence. The feature selection procedures (causal inference, and thresholded spike-and-slab regression, respectively) used to choose the features included in this model are detailed in the subsequent sections. The observed incidence data was the number of confirmed cases of malaria at each health facility each month. This can be thought of as `aggregated' data, in the sense that the cases observed at a given health facility are drawn from the surrounding area, rather than originating in the same location as the health facility. We therefore modelled incidence using disaggregation regression~\citep{wilson2020pointless, taylor2018continuous}, a method designed for aggregated data where the case generating process is modelled on a fine grid of pixels. The likelihood of the observed data is then given by a weighted sum of these processes over the modelled catchment area of each facility.  Another important factor to consider was that a large proportion of the population may not seek treatment for fever within the public healthcare system and therefore the cases observed at health facilities represent only a certain fraction of all malaria cases. Accounting for this treatment-seeking behaviour allows for relationships between covariates and incidence to be learned more reliably and for predictions of true incidence to be made at the pixel level. 

\subsubsection{Treatment-seeking population}
The proportion of the population at a given location that would seek treatment for fever in the formal healthcare system was modelled as a function of the time required to travel to the nearest health facility. For each pixel, the travel time to each health facility was calculated using a friction surface (defining travel time through each pixel; developed by \citealt{weiss2018global}) and a least cost algorithm \citep{dijkstra1959note}. A logistic function was used for this relationship between travel time and treatment-seeking (similar to the functional forms considered by \citealt{alegana2012spatial}), with the proportion who would seek treatment in a pixel set  equal to
$$\frac{\alpha}{1 + \exp(\sigma t)} + \beta; \ (\alpha + \beta \leq 1)$$
where $t$ is the travel time (in minutes) from this pixel to the nearest health facility. Parameters values $\alpha=0.6$, $\sigma=0.00916$ (i.e., $1/\sigma \approx 110$ minutes), $\beta=0.15$ were chosen such that the maximum and minimum possible treatment-seeking proportions were 0.6 and 0.15, and the treatment-seeking proportion at $t=120$ minutes was 0.3. These produced a relationship between treatment-seeking and travel time that was similar to that observed in the 2013 and 2016 Malaria Indicator Surveys~\citep{mis2013, mis2016} (see supplementary material Figure S2) and matched estimated national treatment-seeking rates~\citep{battle2016treatment}.

\subsubsection{Catchment model}
The health facility catchment model was also based on the friction surface-derived travel time estimates. Of individuals in pixel $i$ who seek treatment, the proportion seeking treatment at health facility $j$, $p(\mathrm{pixel}_i\rightarrow\mathrm{HF}_j)$, was supposed zero if the travel time to heath facility $j$ was more than 200 minutes, and otherwise modelled as proportional to the health facility attractiveness (a per-facility parameter) divided by the square of the travel time to that health facility. That is, defining 
    \[\tilde{p}(\mathrm{pixel}_i\rightarrow\mathrm{HF}_j) := \begin{cases} t(\mathrm{pixel}_i\rightarrow\mathrm{HF}_j)^{-2}w_j & \mbox{if } t(\mathrm{pixel}_i\rightarrow\mathrm{HF}_j) \leq 200 \\ 
    0 & \mbox{otherwise} \end{cases} \]
where $t(\mathrm{pixel}_i\rightarrow\mathrm{HF}_j)$ was the travel time from pixel $i$ to health facility $j$ and $w_j$ was the attractiveness of health facility $j$, the proportion seeking treatment at facility $j$ was
\[p(\mathrm{pixel}_i\rightarrow\mathrm{HF}_j) = \frac{\tilde{p}(\mathrm{pixel}_i\rightarrow\mathrm{HF}_j)}{\sum_{k=0}^{N_\textrm{HF}}\tilde{p}(\mathrm{pixel}_i\rightarrow\mathrm{HF}_k)}\]
where $N_\textrm{HF}$ was the total number of health facilities. The inclusion of an attractiveness parameter for each health facility allows representation of the impact of facility-specific latent factors that may affect treatment-seeking preferences. These weightings were learned jointly with the incidence surface within the hierarchical Bayesian structure of the disaggregation regression model.

\subsubsection{Disaggregation regression model} \label{disag_model}
The number of cases occurring in pixel $i$ in month $t$ that would be seen in the healthcare system, $Y_{it}$, was modelled as a Poisson random variable 
\[Y_{it} \sim \textrm{Poisson}(\lambda_{it} \times \mathrm{pop}_i)\]
where $\lambda_{it}$ was the latent incidence rate per person in that pixel in that month and $\mathrm{pop}_i$ was the population in pixel $i$, adjusted for treatment-seeking behaviour. The log incidence rate was modelled as the sum of the effect of environmental and socioeconomic covariates and a spatial Gaussian process,
\[\log\lambda_{it} = f(X_{it}) + \mathrm{GP}(s_i)\]
where $X_{it}$ were covariate values (in pixel $i$ at time $t$), GP was the Gaussian process and $s_i$ was the location of pixel $i$. Two different models were used based on two different functional forms of $f$, as described below. A M\'atern kernel was used for the Gaussian process,  parameterised by the range, $\rho$, and marginal standard deviation, $\sigma$, which were learned. The number of cases observed at health facility $j$ in month $t$, $Y^{\mathrm{HF}}_{jt}$ was then a weighted sum of cases in each pixel, with the weights from the catchment model,
\[Y^\mathrm{HF}_{jt} = \sum_{i=1}^{N_\textrm{pixel}}p(\mathrm{pixel}_i\rightarrow\mathrm{HF}_j) \,Y_{it}\]
where $N_\textrm{pixel}$ was the total number of pixels. In other words, the number of cases observed at health facility $j$ was the sum over all pixels of the proportion of cases occurring in each pixel that would seek treatment at health facility $j$. Assuming that the number of cases in each pixel is independent conditional on the underlying rate $\lambda_{it}$, this sum also follows a Poisson distribution,
\[Y^\mathrm{HF}_{jt} \sim \textrm{Poisson}\left(\sum_{i=1}^{N_\textrm{pixel}}p(\mathrm{pixel}_i\rightarrow\mathrm{HF}_j)\times \lambda_{it}\times \mathrm{pop}_i\right)\]
which gives the likelihood of the response data. The functional forms used for $f(\cdot)$, the additive impact of covariates on the log incidence rate, were as follows.
\begin{enumerate}
    \item{\textit{Linear model.} The covariates were assumed to have a linear effect on $\log$ incidence,
    \[f(X_{it}) = \beta_0 + \beta^TX_{it}\]
    where $\beta_0, \beta$ were parameters to be learned.}
    \item{\textit{Gaussian process model.} Non-linear effects of covariates were considered by modelling the effect of each with a univariate GP,
    \[f(X_{it}) = \beta_0 + \sum_{l=0}^{N_\textrm{cov}}g_l((X_{it})_l) \mathrm{\ with\ }g_l(\cdot) \sim \mathrm{GP}_\lambda \mathrm{\ for\ all\ } l=1,\ldots,N_\mathrm{cov}\]
    where $N_\textrm{cov}$ was the number of covariates. The covariate space GPs were assigned a unit-variance squared exponential kernel to structurally promote smoothness,
    \[k_l(x, y) = \exp\left(-\frac{(x-y)^2}{2\lambda^2}\right)\]
    where the parameter $\lambda$ was common across all variables and was learned during model fitting.
    }
\end{enumerate}
In each case, the model was completed by assigning prior distributions to the various parameters. A penalised complexity prior was used for the range and standard deviation of the M\'atern kernel of the spatial Gaussian process \citep{fuglstad2019constructing}, chosen such that
\begin{align*}
    p(\rho < 1) & = 0.01,\mathrm{\ and\ } \\
    p(\sigma > 0.5) & = 0.01.
\end{align*}
The range parameter is the distance between points where the correlation is approximately 0.1 and a distance of 1 (with respect to coordinates given by longitude and latitude) is equivalent to approximately 110km. The attractiveness weights of the health facilities were given independent $\log$-Normal priors,
\[\log w_j \sim \mathrm{N}(0, 0.25^2)\]
which placed most of the density on values of $w_j$ between $1/\sqrt{3}$ and $\sqrt{3}$, therefore assuming a-priori that it was unlikely for any health facility to be more than 3 times as attractive as any other. The intercept term, $\beta_0$, was given a Normal prior with a mean 0 and standard deviation 2. In the linear model the entries of the vector $\beta$ were given independent Normal priors with mean 0 and standard deviation 1 and in the Gaussian process model a $\log$-Normal prior was applied to $\lambda$ with mean 1.5 and standard deviation 0.1. These priors were chosen to produce prior predictive incidence rates that were on a similar scale to those observed. 

\subsubsection{Forward temporal estimation task}
We assessed the power of the linear and Gaussian process models to forecast malaria incidence under three scenarios: (i) using all available features (referred to as `no selection'); (ii) using only those selected by the causal inference procedure, and (iii) using only those selected using thresholded spike-and-slab regression.  In each case, the incidence model (as described above) was fit using one year of data from training locations (a random subsample of all health facilities) and then forward estimates were made for the monthly incidence rate over the following two years at all health facilities.  Since the model-based estimates for each month depend on covariates from the same or recent months we do not consider these outputs to be, in a strict sense, forecasts (\textit{cf}. \citealt{friedman2020predictive}).  However, the ability to provide forward estimates of incidence to months beyond those for which case data are yet available is nevertheless of value to malaria control programs.  Covariate data are often ready for use many months earlier than case data for which a lead time is incurred through the stages of collection, processing, validation, and aggregation in many malaria endemic settings.

The number of training locations was varied, as was the starting month of the three year time period studied (which we refer to as iterations). For example, in the first iteration the model was fit to data from January 2013 to December 2013 and forward estimates were made from January 2014 to December 2015, and in the second iteration the model was fit using data from February 2013 to January 2014 and forward estimates were made from February 2014 to January 2016. With four years of data in total, there were therefore 13 iterations.

The main metric of performance used was correlation between observed and predicted incidence rates. Correlation is often the preferred metric when evaluating malaria risk mapping, as relative spatial and temporal trends are of most interest to program managers. For example, the allocation priority of limited resources or targeting of interventions may be based on the relative estimated risk of different regions. A model which accurately distinguishes lower and higher risk areas can therefore be of more use than a model which has smaller absolute error but is less accurate at establishing relative rankings. We considered overall correlation between predicted and observed rates (over all locations and months) and `temporal correlation', which we defined as the correlation between the predicted and observed time series (over the 24 months forecasted) at each location, averaged over all locations. Accurately estimating temporal patterns in transmission is useful for interventions that are based around high transmission periods (such as seasonal malaria chemoprevention and indoor residual spraying). We also considered this temporal correlation limited to only the second year of predictions (months 13-24) in order to investigate model performance further into the future. Predictions were also assessed in terms of root mean square error (RMSE).

\subsection{Causal selection}\label{sec:causalselection}

\subsubsection{Spatiotemporal prewhitening}\label{subsec:prewhiten}
The first step of the causal feature selection procedure is a prewhitening step to remove large-scale spatial and temporal trends from all variables. The idea here is that the observed data is likely to have contamination from spatiotemporal fluctuations not captured by the available feature set, which can both violate the assumption of independent and identically distributed (iid) observations that underlie many independence tests and act as a confounder, inducing correlations between variables where no causal relationship exists. We therefore fit an autoregressive model to each variable to remove these general spatiotemporal trends, resulting in approximately iid residuals, a procedure formalised in the context of causal discovery algorithms by \cite{flaxman2016gaussian}. The model used for the incidence data was a spatiotemporal version of the model described in Section \ref{subsec:incidence_model} which used no covariates. That is, the model was the same as described in this section except that the incidence rate in pixel $i$ at time $t$ was modelled as,
\[\log\lambda_{it} = \mathrm{GP}(s_i, t)\]
where $s_i$ was the location of pixel $i$, where GP was a separable Gaussian process with M\'atern structure in the spatial component and an AR1 structure over time. The residuals for each health facility were then calculated by subtracting the rate in the fitted model from the observed rate. Health facility attractiveness weights were learned as part of this catchment model and the resulting proportions of the population seeking treatment at each facility were used in the aggregated independence tests (see Section~\ref{subsec:aggregate_indep}). In order to enforce that this prewhitening model learned broad spatial patterns, the range and scale parameters of the M\'atern covariance were given log-normal priors with a mean of $\exp(2)$ and $\exp(-2)$, respectively, both with a standard deviation (on the log sale) 0.1. The environmental and socioeconomic covariates were modelled with a spatiotemporal Gaussian process of this form directly. For a temporal covariate, the value in pixel $i$ at time $t$, $Y_{it}$, was modelled as,
\[Y_{it} = \mathrm{GP}(s_i, t)\]
where again the Gaussian process was the product of a M\'atern process over space and AR1 process over time. For a spatial covariate (such as elevation) the value in pixel $i$, $Y_i$, was modelled as,
\[Y_i = \mathrm{GP}(s_i)\]
with a M\'atern covariance. In both cases, the same log-Normal priors were used as in the prewhitening model for incidence. 

For the prewhitening of both the incidence data and covariates a strong \textit{a priori} degree of regularisation is encoded in these priors, with the intention being to limit exposure to overfitting in this step. Therefore it is possible that less spatiotemporal structure was removed than would be in the ideal prewhitening procedure.  Note, however, that residual spatiotemporal structure is expected (on average) to result in a causal discovery algorithm incorrectly retaining edges, which is less problematic for feature selection than incorrectly removing them. 

\subsubsection{Kernel-based independence tests}
A number of independence and conditional independence tests have been proposed based on the theory of reproducing kernel Hilbert spaces (RKHS)~\citep{gretton2005kernel, gretton2005measuring, gretton2008kernel, jitkrittum2017adaptive, zhang2012kernel, strobl2019approximate}. The general approach of these tests is to define covariance and cross-covariance operators in an RKHS which encode information about (conditional) dependence, and to then derive appropriate test statistics based on the expected properties of these operators. In this analysis we used the Randomized Conditional Independence Criterion proposed by~\cite{strobl2019approximate}, however any kernel-based test could be used in its place within this causal discovery procedure. In particular, our proposed method for including variables observed at different spatial and temporal resolutions works by modifying the kernel function, and therefore is not specific to any particular test. 

For a set $X$, a Hilbert space $\mathcal{H}_X$ of functions $X\rightarrow\mathbb{R}$ is a RKHS if the evaluation functional
\begin{align*}
    \delta_x:&\,\mathcal{H}_X\rightarrow\mathbb{R} \\
    &\,f \mapsto f(x)
\end{align*}
is a bounded linear functional for any $x\in X$. In this case, there exists a unique kernel function $k:X\times X\rightarrow\mathbb{R}$ associated with $\mathcal{H}_X$ such that $k(\cdot, x)\in\mathcal{H}_X$ for all $x\in X$ and $k$ satisfies the reproducing property,
\[\langle f,\,k(\cdot, x)\rangle_{\mathcal{H}_X} = f(x)\]
for all $x\in X$, $f\in\mathcal{H}_X$. The map $\phi:x\mapsto k(\cdot, x)$ is the feature map of $\mathcal{H}_X$. This feature map is employed in many `kernelised' machine learning algorithms, such as kernel ridge regression or kernel vector support machines, which use these feature maps (whether implicitly or explicitly) to embed data points into high dimensional spaces. 

To define the covariance and cross covariance operators used in independence testing, consider random variables, $X, Y$, with marginal probability distributions, $P_X, P_Y$, and joint distribution, $P_{XY}$. Let the co-domains be $\mathcal{X}, \mathcal{Y}$ and let $\mathcal{H}_X:\mathcal{X}\rightarrow\mathbb{R}$ and $\mathcal{H}_Y:\mathcal{Y}\rightarrow\mathbb{R}$ be separable RKHSs (such that an orthonormal basis exists for each) with corresponding kernels, $k_X, k_Y$, and feature maps, $\phi_X, \phi_Y$. The mean embedding of $X$ can then be defined as
\[\mu_X:= E_X[\phi_X(X)] = \int_\mathcal{X}k(\cdot, x)\,\mathrm{d}P(x)\]
which exists and is an element of $\mathcal{H}_X$ when $E_{XX'}[k(X, X')] < \infty$. Suppose $E_{XX'}[k(X, X')] < \infty$ and $E_{YY'}[k(Y, Y')] < \infty$ and let $\mu_Y\in\mathcal{H_Y}$ be the mean embedding of Y. The associated cross-covariance operator is then defined as the linear operator, $\Sigma_{XY}:\mathcal{H}_Y\rightarrow\mathcal{H}_X$, such that
\begin{align*}
    \langle f, \Sigma_{XY}g\rangle_{\mathcal{H}_X} &= E_{XY}[f(X)g(Y)] - \mu_X\mu_Y \\
    &= E_{XY}[f(X)g(Y)] - E_X[f(X)]E_Y[g(Y)] \\
    &= \mathrm{Cov}[f(X), g(X)]
\end{align*}
for any $f\in\mathcal{H}_X$ and $g\in\mathcal{H}_Y$~\citep{baker1973joint, fukumizu2004dimensionality}. ~\cite{gretton2005measuring} show that for a suitable choice of kernels, $k_X, k_Y$, and compact $\mathcal{X}, \mathcal{Y}$, the Hilbert-Schmidt norm of the cross-covariance operator is zero if and only if $X$ and $Y$ are independent. This is the foundation of a number of independence tests, including including the Hilbert Schmidt Independence Criterion (HSIC)~\citep{gretton2005kernel, gretton2005measuring} and the Finite Set Independence Criterion~\citep{jitkrittum2017adaptive}. Similarly, with three random variables, $X, Y, Z$, the operator, $\Sigma_{XY|Z}:\mathcal{H}_Y\rightarrow\mathcal{H}_X$, can be defined as,
\[\Sigma_{XY|Z} = \Sigma_{XY} - \Sigma_{XZ}\Sigma^{-1}_{ZZ}\Sigma_{ZY}\]
which is the partial conditional cross-covariance operator of $X, Y$ given $Z$~\citep{fukumizu2004dimensionality, strobl2019approximate}. This operator is such that
\[\langle g, \Sigma_{XY|Z}f\rangle_{\mathcal{H}_X} = E_Z[\mathrm{Cov}_{XY|Z}(f(X), g(Y)|Z)]\]
for all $f\in\mathcal{H}_X$ and $g\in\mathcal{H}_Y$ \citep[Proposition 5]{fukumizu2004dimensionality}. Furthermore, defining $\dot{X} = (X, Z)$ it can be shown under certain conditions~\citep{fukumizu2004dimensionality} that
\[\Sigma_{\dot{X}Y|Z} = 0 \Leftrightarrow \indep X  Y |Z.\]
This operator is central to a number of conditional independence tests, including the Kernel Conditional Independence Test (KCIT)~\citep{scholkopf2000kernel}. Here we adapted the Randomized Conditional Independence Test (RCIT)~\citep{strobl2019approximate}, an approximation of KCIT which utilizes random Fourier features in order to scale linearly with sample size. In simulations carried out by \cite{strobl2019approximate}, RCIT and KCIT have comparable accuracy.



\subsubsection{Testing for independence with aggregated data}\label{subsec:aggregate_indep}
The variables included in this analysis are on different spatial scales, as the case data are aggregations over catchment areas while environmental and socioeconomic variables are available on a fine-scale grid. This presents a challenge to causal inference, as relationships learned between variables on an aggregate level do not necessarily hold at individual level, which is known as the ecological fallacy~\citep{wakefield2006health}. This means that variables that are dependent at the individual level may appear independent when aggregated, and vice versa. As the true disease mechanism occurs at the individual level, causal inference may be expected to be most effective when based on relationships learned at (or near) this scale. To address this problem, we propose a distributional approach, similar to distribution regression methods for ecological inference~\citep{szabo2015two, gartner2002multi, flaxman2015supported}. 

Suppose we have observations, $y_1,....,y_n$, of random variable, $Y$, and corresponding groups of observations of $X$, $\{x_1^i\}_{i=1}^{N_1},...,\{x_n^i\}_{i=1}^{N_n}$, and we wish to test $X$ and $Y$ for independence. For example, $y_j$ could be the observed incidence rate at a health facility in a specific month and $\{x_j^i\}_{i=1}^{N_j}$ could be the temperature in that month in each pixel within the catchment area of the health facility. Given a RKHS, $\mathcal{H}_X$, on $X$ (with kernel, $k_X$, and feature map, $\phi_X$) we can embed each group of observations of $X$ into a single point in this space
\[\{x_j^i\}_{i=1}^{N_j} \longmapsto \hat{\mu}_j := \sum_{i=1}^{N_j}w_j^i\phi_X(x_j^i).\]
The weights $w_j^i$ are context-specific and in our case we weighted $\phi_X(x_j^i)$ by the number of people in the corresponding pixel who would seek treatment at health facility $j$ (these proportions were modelled in the prewhitening step described in Section~\ref{subsec:prewhiten}). Therefore this weighted sum over the catchment area is equivalent to an unweighted average over the population expected to attend this facility. As the notation suggests, $\hat{\mu}_j$ could be thought of as an empirical estimate of the mean embedding of some distribution associated with the variable $X$ in this catchment area $j$. Unlike many other transformations of the grouped observations of $X$ (such as the mean), with a suitable choice of kernel this embedding preserves all information from the individual observations. Our problem is then re-framed as testing for independence between $Y$ and this aggregate random variable, which we denote $X_\mu$, using joint observations, $(y_1, \hat{\mu}_1),...,(y_n, \hat{\mu}_n)$, against which any kernel-based independence test can be applied. A number of kernels could be used for $X_\mu$ (see \citealt{szabo2015two}) but as the values of $X$ have already been embedded in a Hilbert space, we used the simplest choice, the linear kernel
\[k_\mu(\hat{\mu}_i, \hat{\mu}_j) = \langle\hat{\mu}_i, \hat{\mu}_j\rangle_{\mathcal{H}_X}\]
which can be written in terms of the kernel, $k_X$,
\[k_\mu(\hat{\mu}_i, \hat{\mu}_j) = \sum_{k=1}^{N_i}\sum_{l=1}^{N_j}w_i^kw_j^lk_X(x_i^k, x_j^l).\]
This idea extends naturally to conditional independence tests. If we wish to test $X$ and $Y$ for independence conditional on variables $\{Z_1,...,Z_k\}$ then any of the variables $Z_i$ which are available at the fine scale may be embedded into a RKHS in the same way as $X$, while any that are aggregated can remain unchanged. 

Furthermore, many scalable kernel independence tests make use of random Fourier features (RFFs; such as the RCIT and approximations to the HSIC), to make a low rank approximation to high (often infinite) dimensional RKHSs. We can find similar approximations to $k_{\mu}$ as follows. The RFF method uses vectors of the form
\[z(x) = (\cos(v_1^Tx),\sin(v_1^Tx),...,\cos(v_D^Tx),\sin(v_D^Tx))/\sqrt{D}\]
to approximate a kernel $k_X$,
\[k(x, y) \approx z(x)^Tz(y)\]
where the values $v_1,...,v_D$ are randomly sampled from a distribution that depends on the choice of kernel $k_X$. A suitable distribution exists for real valued translation invariant positive definite kernels on $\mathbb{R}^d$ by Bochner's theorem. The RKHS that corresponds to the feature map $z$ is $2D$-dimensional and many kernel methods can be reformulated to take advantage of this low dimensionality. Gaussian process regression, for example, involves solving linear systems of dimension equal to the number of observations and therefore scales poorly in high data settings, but can be rewritten to in terms of $2D$-dimensional systems. To approximate $k_\mu$, we note that
\begin{align*}
   k_\mu(\hat{\mu_i}, \hat{\mu_j}) &= \sum_{k=1}^{N_i}\sum_{l=1}^{N_j}w_i^kw_j^lk_X(x_i^k, x_j^l) \\
   &\approx \sum_{k=1}^{N_i}\sum_{l=1}^{N_j}w_i^kw_j^lz(x_i^k)^Tz(x_j^l) =  \left(\sum_{k=1}^{N_i}w_i^kz(x_i^k)\right)^T\left(\sum_{l=1}^{N_j}w_j^lz(x_j^l)\right).
\end{align*}
Therefore, the map 
\[z_{\mu} : \{x_j^i\}_{i=1}^{N_j}  \longmapsto \sum_{l=1}^{N_j}w_j^lz(x_j^l)\]
can be used to approximate $k_{\mu}$ and allow us to work in a low dimensional RKHS, just as with the map $z$ above. In other words, the weighted sum of the features $z(x_j^l)$ for the individual observations (with respect to $k_X$) are suitable features for approximating $k_\mu$. Computing the RCIT test statistic involves a random Fourier feature approximation of kernel ridge regression. In the aggregate case, this corresponds to an approximation of distribution regression, as carried out by \cite{flaxman2015supported}.

\subsubsection{The PC algorithm}
As discussed in Section \ref{intro:causal_discovery}, by the Causal Markov Condition and Faithfulness any DAG implies a set of conditional independence relationships in the corresponding joint distribution over all variables. Constraint-based causal discovery algorithms use the conditional independence relationships inferred from the observed data to constrain the set of possible DAGs that could have generated this data. However, the set of possible DAGs is often large (growing exponentially with the number of variables), so an efficient strategy is needed to find the set of compatible DAGs. 

The PC algorithm~\citep{spirtes1995learning, spirtes2000causation} is one of the most well-known algorithms for constraint-based causal discovery and is efficient, running in polynomial time (as a function of the number of variables) when the true underlying DAG is sparse. The first stage of the algorithm begins with a complete undirected graph and uses conditional independence relationships to recursively remove edges. Letting $S$ be the set of random variables, the intuition behind this is that for any two variables, $X,Y\in S$, if the true DAG contains an edge $X\rightarrow Y$ or $Y\rightarrow X$ then (due to Faithfulness) these variables will be dependent unconditionally and when conditioning on any subset of $S\backslash\{X, Y\}$. Therefore we could test for an edge between $X$ and $Y$ by testing $X, Y$ for independence unconditionally and conditional on all subsets of $S\backslash\{X, Y\}$, removing the edge if $X$ and $Y$ are ever found to be independent. However, this procedure scales exponentially with the number of variables due to the number of conditioning subsets (even if the true DAG is sparse). The PC algorithm makes use of the fact that if there does not exist an edge between $X, Y$, then $X$ and $Y$ are independent conditional on some subset of variables adjacent to $X$ or $Y$. This reduces the number of conditioning sets substantially when the true DAG is sparse. 

The output of the first stage of the algorithm is an undirected graph (or `skeleton'). The second stage of the output attempts to direct these edges, where possible. This is primarily done by identifying `v-structures', triples $X,Y,Z\in S$ such that $X\rightarrow Y \leftarrow Z$ and $X, Z$ are not connected. With this structure, $X$ and $Z$ (which are unconditionally independent) are dependent given $Y$. This phenomenon, where conditioning on a common cause induces a dependence between variables, is often called `collider bias' (an example of which is selection bias). Therefore, we can test if an undirected triple $X\,\hbox{---}\,Y\,\hbox{---}\,Z$ is a v-structure by testing $X,Z$ for independence given $Y$. The second stage of the PC algorithm uses similar logic to identify v-structures within the whole skeleton structure based on the independence tests performed in the first stage. Additional edges can be directed by ruling out directions that would produce cycles or v-structures that were not identified by the independence tests. The PC algorithm assumes that there are no unmeasured common causes for observed variables (this assumption is known as causal sufficiency). We suggest that this is satisfied with respect to the environmental variables used. The two human factors considered, accessibility to cities and presence of stable nighttime lights could potentially have common causes, as both are related to urban development. However, we would in fact argue that urban development (measured by the presence of nighttime lights) is the primary cause of accessibility to cities and most, if not all, other causes of accessibility act indirectly through development.

To perform causal discovery on our dataset, we ran the PC algorithm twice, in both cases using RCIT to carry out the independence tests, with the aggregated version (as described in Section~\ref{subsec:aggregate_indep}) used as necessary. First, we ran the algorithm without including the malaria incidence variable. This is possible because incidence was assumed to not to cause any other variables, and therefore valid causal discovery for the remaining variables could take place without it. The benefit of this first step is that observations are not limited to the locations and times where the response data is available. We then used the resulting partially directed DAG and added in the incidence variable, including edges between incidence and every other variable. We then ran a modified version of the PC algorithm, in which only edges involving incidence were tested. A Gaussian kernel was used within the independence test,
\[k(x, y) = \exp\left(-\frac{(x-y)^2}{\lambda^2}\right)\]
with bandwidth, $\lambda$, chosen to be the median pairwise distance between observations. This heuristic has been used extensively in the application of kernel methods and has been shown empirically to lead to good performance~\citep{zhang2012kernel, gretton2008kernel, strobl2019approximate, garreau2017large}. The KCIT test has also been shown to be robust to variation in bandwidth parameter around this value~\citep{zhang2012kernel}.

We repeatedly sub-sampled the data and ran the above two-step method to produce a number of output graphs. The confidence that a given feature was a direct cause of incidence was then quantified by the proportion of graphs in which it was a direct parent of incidence. We note that this proportion was simply a relative ranking of features, rather than representing the probability that a feature was a direct cause. Finally, we selected the top four static features and top four dynamic features to make up the final feature set. This was repeated for each iteration. This bootstrapping of the data is similar to the suggestion by~\citet{austin2004bootstrap} of applying backwards elimination selection to bootstrapped samples in order to improve predictive performance. 
 



% ...

\subsection{Spike-and-slab regression}
Spike-and-slab regression is a sparsity-promoting Bayesian modelling technique originally designed for linear regression~\citep{mitchell1988bayesian, george1993variable, kuo1998variable, ishwaran2005spike}. The name refers to the characteristic mixture model prior applied to the regression coefficients, made up of a point mass at zero (the `spike') and a wide, flat distribution (the `slab'). Spike-and-slab models are advantageous for variable selection as they result in selective shrinkage, shrinking small coefficients towards zero while leaving large coefficients relatively unchanged, in a similar way to the LASSO in the frequentist context~\citep{tibshirani1996regression}. In our implementation we use a continuous bimodal prior for the variance of the prior on the coefficients, rather than a mixture prior on the coefficients, to achieve the same effect, as suggested by \cite{ishwaran2005spike}. In the case of the linear model (as defined in Section~\ref{disag_model}), the only difference between the final regression model and the spike-and-slab model was the prior placed on the regression coefficients. The prior hierarchy used was as formulated by \cite{ishwaran2005spike},
\begin{align*}
    \qquad\qquad\qquad\qquad\beta_l | \phi_l, \tau_l^2 &\stackrel{\mathrm{ind}}{\sim} \mathrm{N}(0, \phi_l\tau_l^2)  \qquad\qquad l=1,...,N_\textrm{cov} \\
    \phi_l|\nu_0, w &\stackrel{\mathrm{iid}}{\sim} (1-w)\delta_{\nu_0}(\cdot) + w\delta_1(\cdot) \\
    \tau_l^{-2}|a_1, a_2 &\stackrel{\mathrm{iid}}{\sim} \mathrm{Gamma}(a_1, a_2) \\
    w &\sim \mathrm{Uniform}[0, 1]
\end{align*}
with $a_1=40, a_2=0.025$ and $\nu_0=0.01$ to produce a bimodal prior for $\phi_l\tau_l^2$. For the Gaussian process model, a modified form of $f$ was used,
    \[f(X_{it}) = \beta_0 + \sum_{l=0}^{N_\textrm{cov}}\beta_l\mathrm{GP}_l((X_{it})_l)\]
with the same prior hierarchy described above. The posterior was sampled using MCMC, with four chains for each repeat with a burn-in of 500 and 3000 steps, to obtain inclusion probabilities for each variable. In a strict sense the spike-and-slab model does not produce a hard selection of variables unless an \textit{a posteriori} processing is applied to summarise the posterior in this manner.  In settings where there are many features to choose from it has been shown that selection by marginal inclusion probability threshold can outperform selection by highest posterior mass \citep{barbieri2018median}. For optimal comparability with  the causal feature selection procedure, we define our thresholded spike-and-slab selection as a restriction to the four static and four dynamic covariates with highest marginal inclusion probabilities respectively. We term this procedure `spike-and-slab selection'.

\section{Results}
\subsection{Feature importance}
\begin{figure}[!htb]
    \centering
    \includegraphics[width=0.9\textwidth
    ]{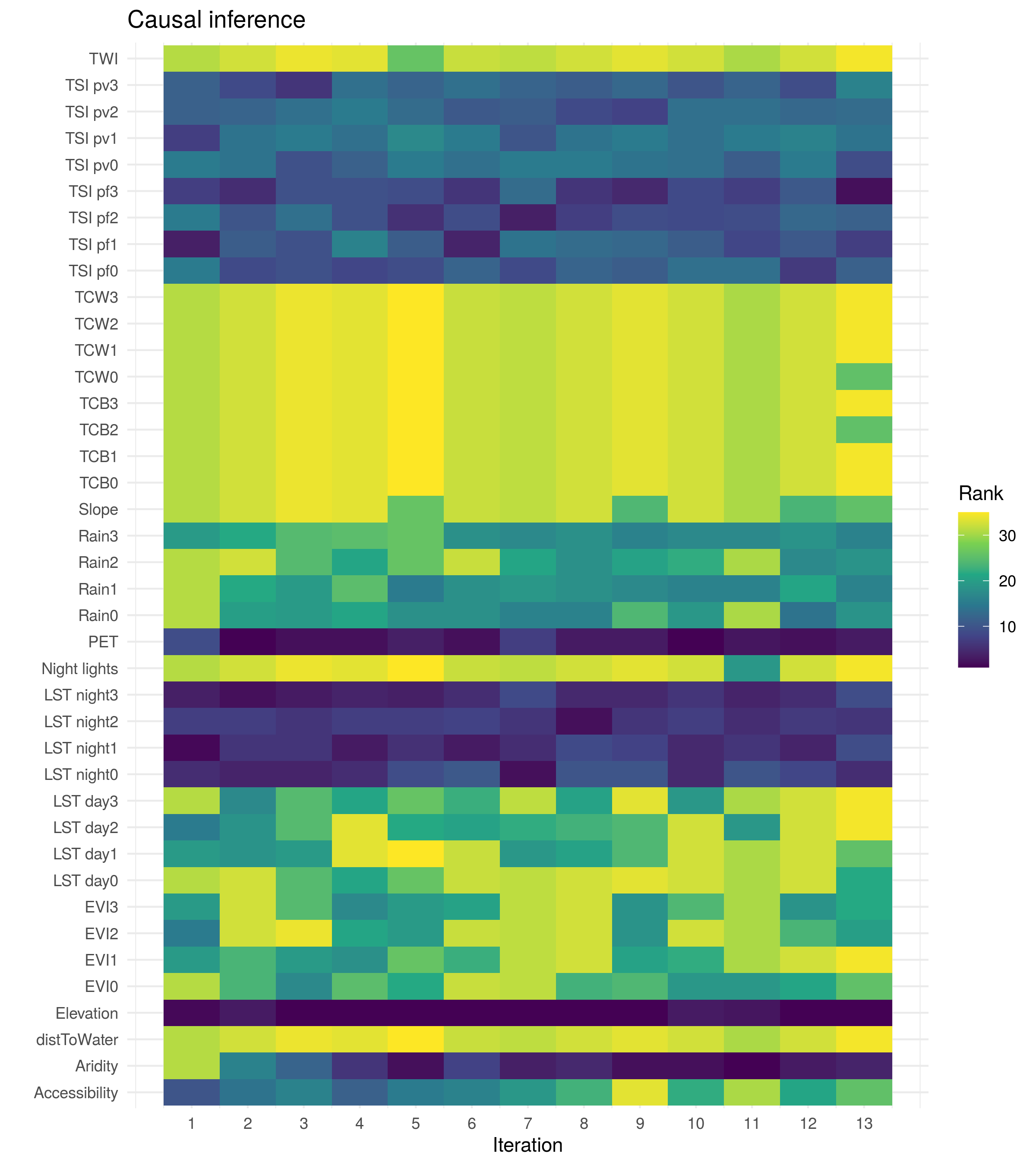}
    \caption{Feature rankings by importance (quantified by the proportion of repeats in which that variable was a direct cause of incidence) according to the causal inference approach, across each of the thirteen data iterations.}
    \label{fig:causal_importance}
\end{figure}

\begin{figure}[!htb]
    \centering
    \includegraphics[width=0.9\textwidth
    ]{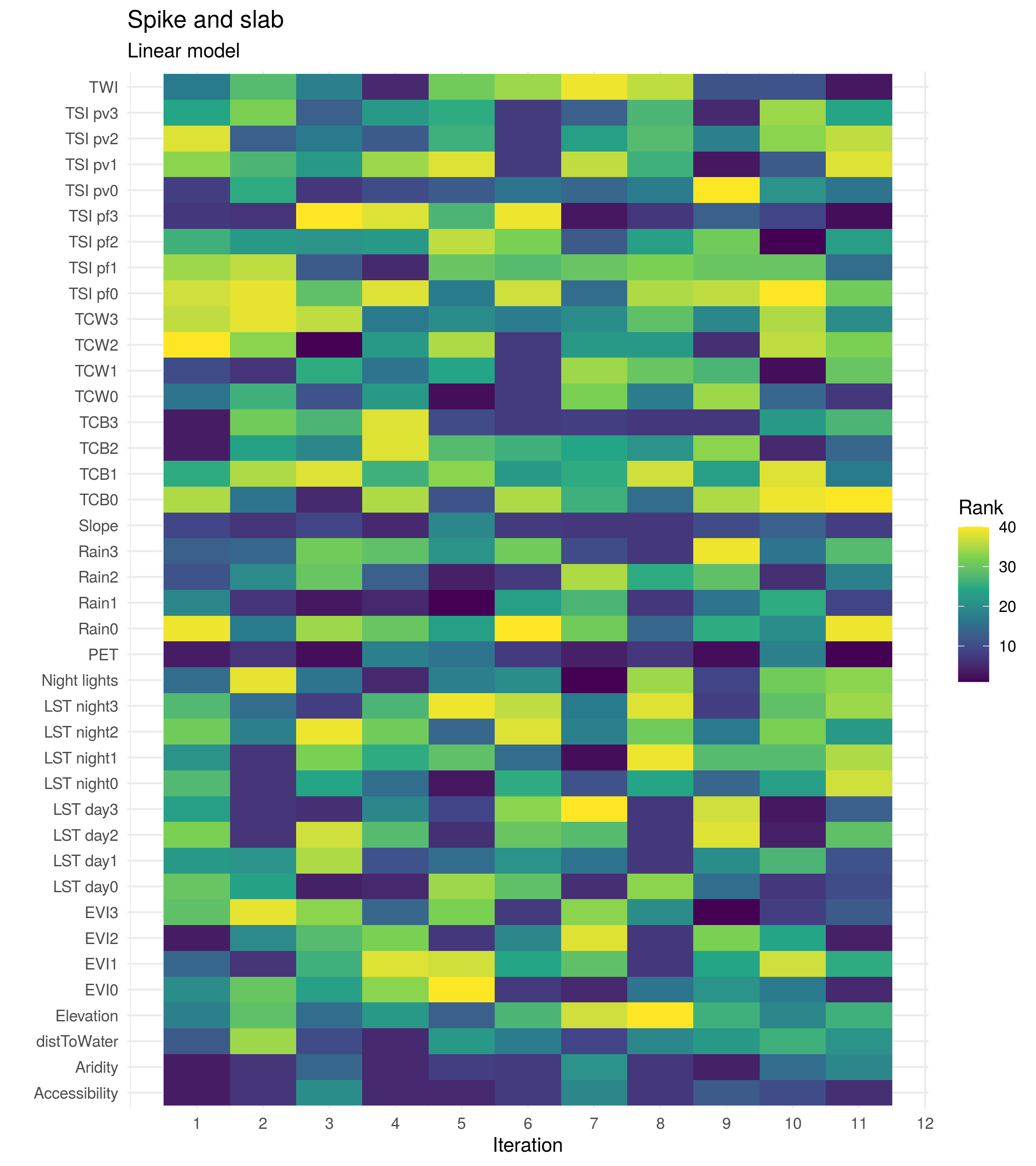}
    \caption{Feature rankings by importance (quantified by the marginal posterior inclusion probability) according to the linear spike-and-slab model, across each of the thirteen data iterations.}
    \label{fig:spike_importance_linear}
\end{figure}

\begin{figure}[!htb]
    \centering
    \includegraphics[width=0.9\textwidth
    ]{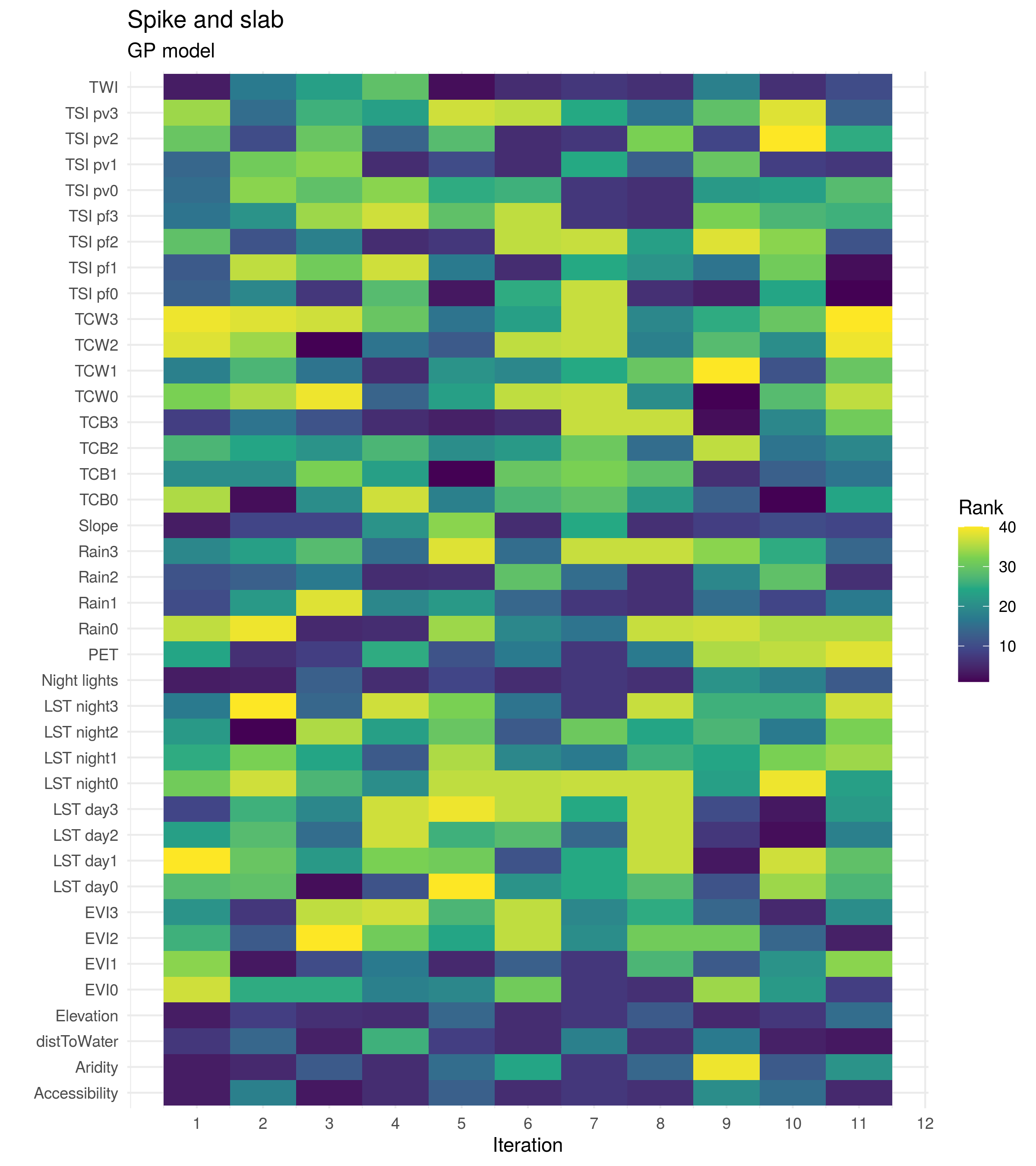}
    \caption{Feature rankings by importance (quantified by the marginal posterior inclusion probability) according to the GP spike-and-slab model, across each of the thirteen data iterations.}
    \label{fig:spike_importance}
\end{figure}

The importance rankings of features returned by the causal inference approach, and by the linear and Gaussian process spike-and-slab models, are presented in Figures \ref{fig:causal_importance}, \ref{fig:spike_importance_linear} and \ref{fig:spike_importance}, respectively.  A total of thirteen rankings of the spatial and spatiotemporal data products available for mapping malaria incidence in Madagascar are shown for each, corresponding to the thirteen overlapping windows of data created for testing purposes by partitioning around different starting months (referred to here as `iterations').  In the case of the causal inference approach, the importance of a feature was quantified by the proportion of outputs in which this feature was a direct cause of incidence from repeated applications of the PC algorithm under the approximate (and noisy) kernel-based independence tests described in Section \ref{sec:causalselection}.  For each of the spike-and-slab models, on the other hand, the importance of a feature was quantified by its marginal posterior inclusion probability.  Since the causal feature selection procedure is model-free, these rankings were used to select variables for both the linear and GP models applied subsequently; whereas feature selection under the spike-and-slab approach is intrinsically tied to the corresponding regression model. 

A comparison of Figure \ref{fig:causal_importance} against Figures \ref{fig:spike_importance_linear} and \ref{fig:spike_importance} reveals a striking contrast between these approaches: feature selection under the causal method is highly stable across the iterations, while that for  the spike-and-slab is highly unstable.  In the former, three static features (elevation, aridity and PET) were consistently high ranked, as were all lags of the LST night and TSI dynamic features (both \textit{Pf} and \textit{Pv} versions).  In the linear spike-and-slab model, four static features (accessibility, aridity, PET and slope) were somewhat consistently high ranked, while the ranking of all dynamic features varied greatly between iterations.  In the GP spike-and-slab model, all static covariates were somewhat consistently high ranked, and again the dynamic features showed great variation.  The between-iteration correlation coefficients for the rankings returned by each approach were 0.87 (causal), 0.13 (linear spike-and-slab) and 0.21 (GP spike-and-slab).  

\subsection{Model performance}
\begin{figure}[!htb]
    \centering
    \includegraphics[width=0.9\textwidth
    ]{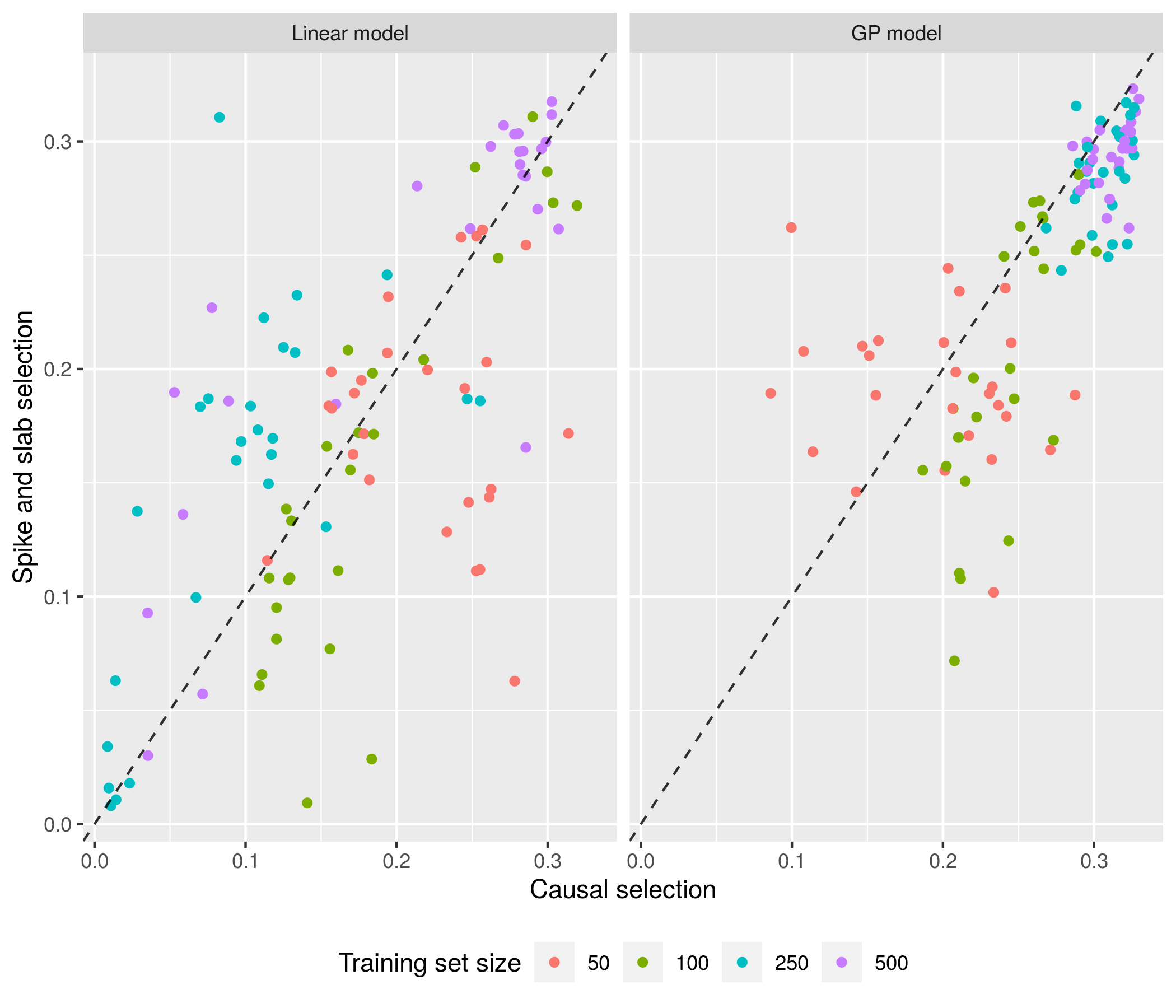}
    \caption{Comparison of correlation between predicted and true incidence rates for the linear (left) and GP (right) models when using spike-and-slab and causal selection.}
    \label{fig:overall_spike}
\end{figure}

\begin{figure}[!htb]
    \centering
    \includegraphics[width=0.9\textwidth
    ]{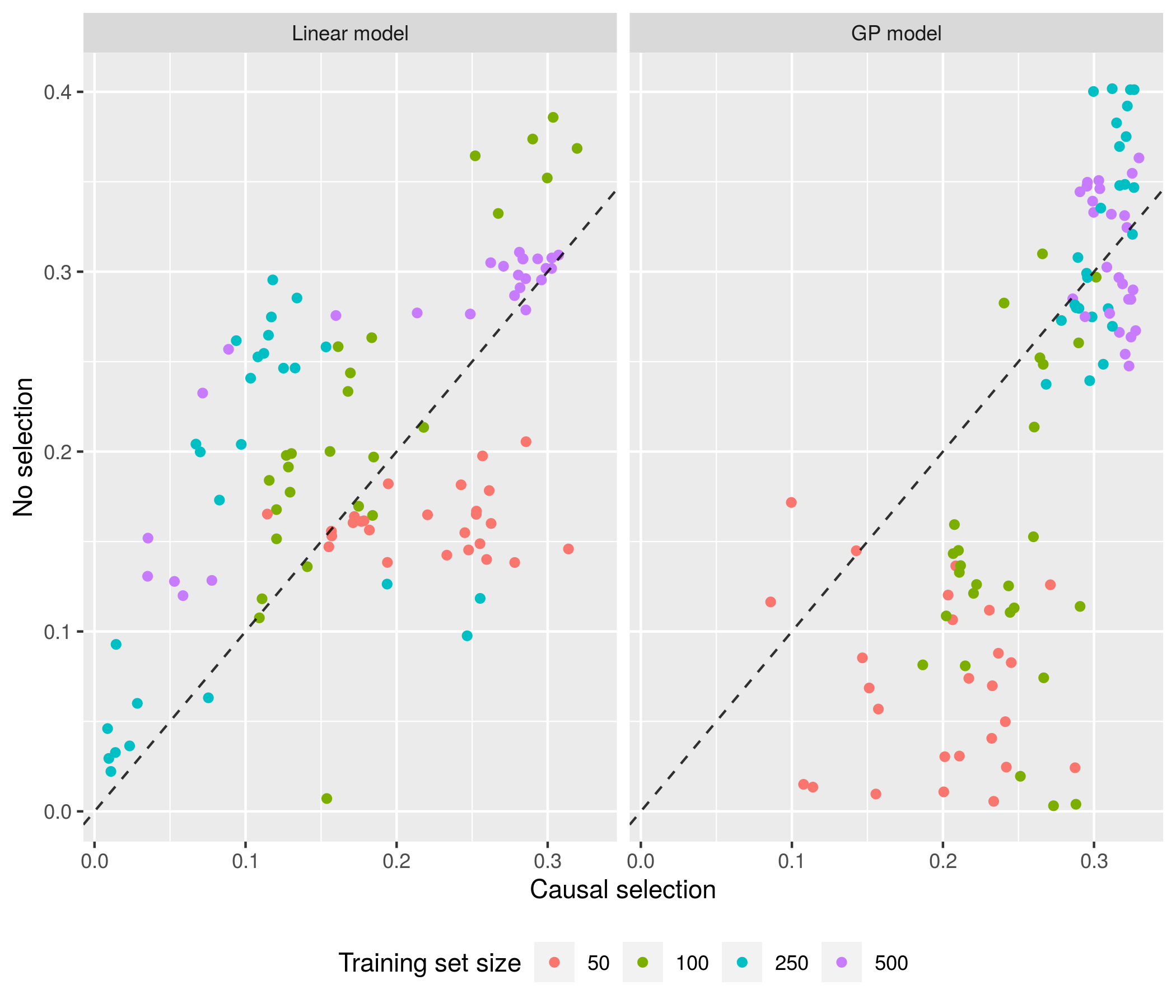}
    \caption{Comparison of correlation between predicted and true incidence rates for linear (left) and GP (right) models when using all features (no selection) and causal selection.}
    \label{fig:overall_all}
\end{figure}

The correlations between true and predicted incidence rates (across all health facilities and months) when using causal and spike-and-slab selection alternately are shown in Figure~\ref{fig:overall_spike}.  For the linear model, overall there was no clear advantage for either selection method but there is a trend of better performance with causal selection for smaller training sets and with spike-and-slab selection for larger training sets. Under the GP model, however, model performance was superior with causal selection in almost all cases except under the smallest training set.  The model performances according to this metric under the alternative of no feature selection is compared with that under causal feature selection in Figure~\ref{fig:overall_all}.  The linear model usually performed better with no selection, except for the smallest size of training set where correlation was consistently higher with causal selection. For the GP model, performance was improved when using causal selection in the large majority of cases, with this effect being clearest for smaller training sets.

\begin{figure}[!htb]
    \centering
    \includegraphics[width=0.9\textwidth
    ]{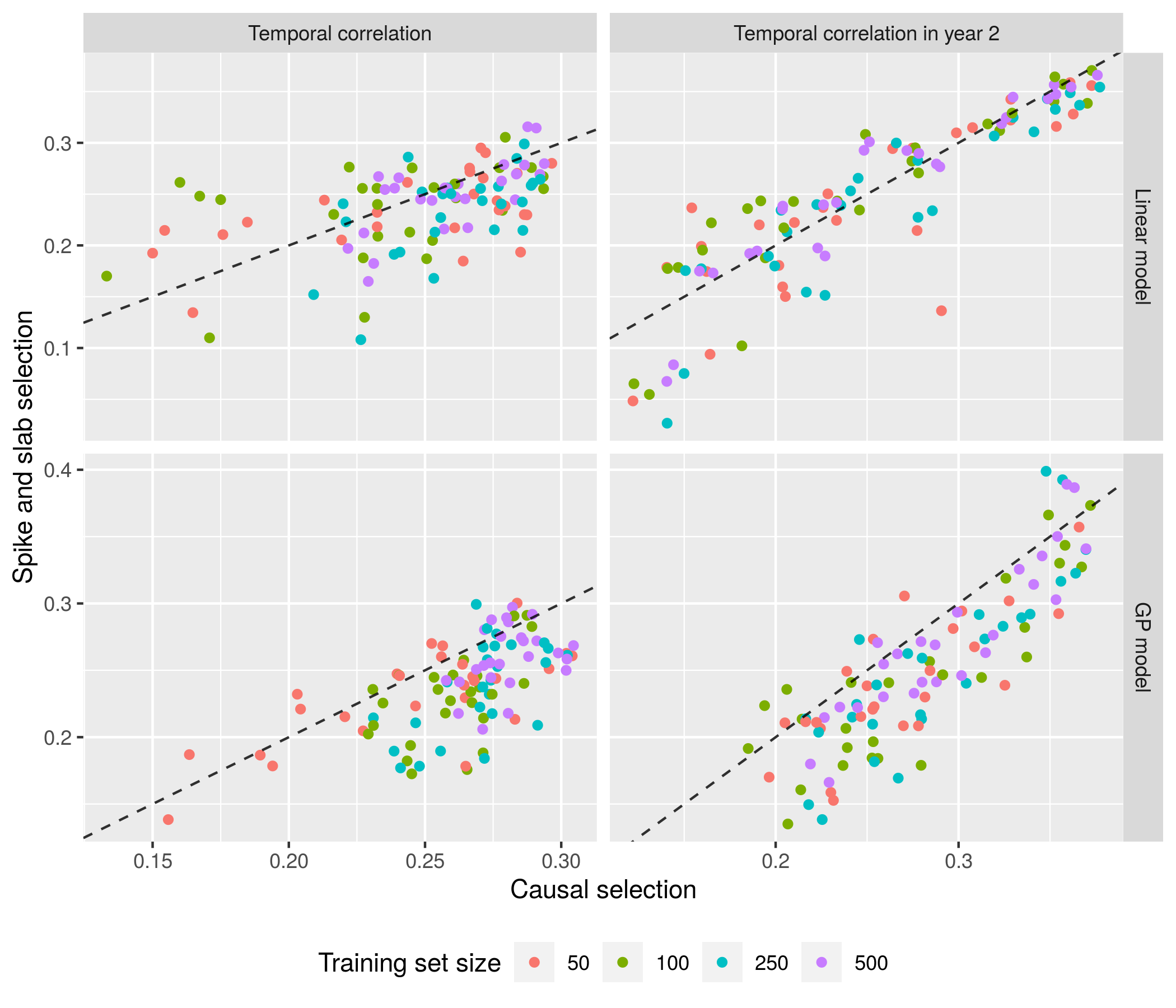}
    \caption{Comparison of temporal correlation for the linear (top) and GP (bottom) models when using spike-and-slab and causal selection. The time-series considered are all 24 months forecasted (left) and months 13-24 (right).}
    \label{fig:location_spike}
\end{figure}

Figure~\ref{fig:location_spike} shows the comparison between model performance in terms of temporal correlation when using causal and spike-and-slab selection. Recall that we define temporal correlation as the correlation between forecasted and observed time series at each health facility averaged over all facilities. This measures the ability of the model to predict temporal trends at each location. As well as calculating the temporal correlation over the entire 24 months forecasted, we also calculated this metric under restriction to months 13-24 in order to emphasise long-term predictive performance.  Overall, both models appear to perform better in terms of temporal correlation when using causal selection compared to spike-and-slab selection. For the linear model, causal selection appeared to improve temporal correlation in the majority of the cases. When this was restricted to just the second year of predictions, the performance using causal or spike-and-slab selection was more similar. The performance of the GP model in terms of temporal correlation was superior using causal selection in almost all cases, both over the whole time frame and in the second year.

\begin{figure}[!htb]
    \centering
    \includegraphics[width=0.9\textwidth
    ]{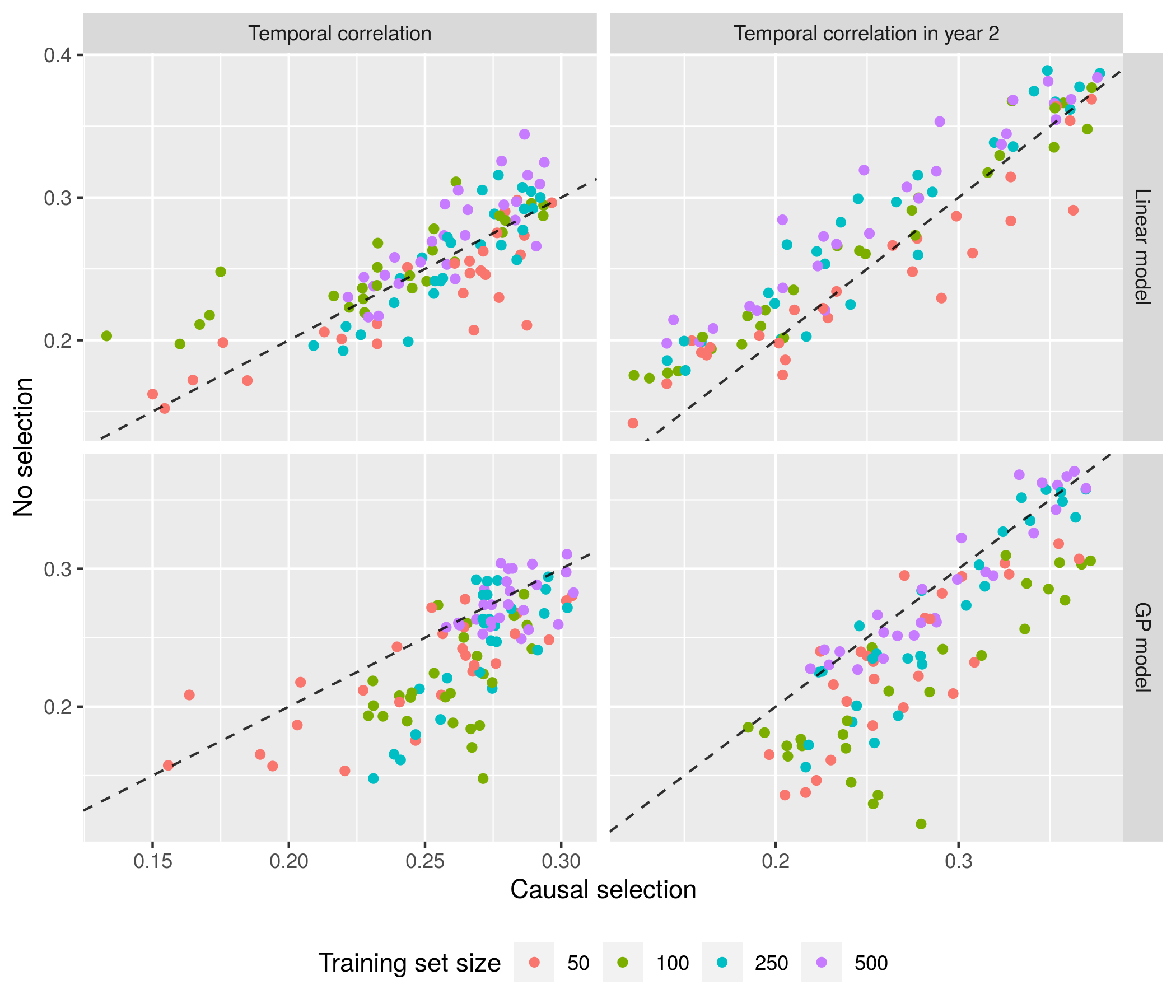}
    \caption{Comparison of temporal correlation for both models when using all features (no selection) and causal selection. The time-series considered are all 24 months forecasted (left) and months 13-24 (right).}
    \label{fig:location_all}
\end{figure}

Causal selection is compared to performing no feature selection in terms of temporal correlation in Figure~\ref{fig:location_all}. For the linear model, results were more favourable (over both time frames) to including all features in the model. The exception to this was for the smallest training sets, in which case temporal correlation was usually higher using causal selection. For the GP model, there was clear advantage to using causal selection over no selection, over both years and in the second year alone. This improved performance was most notable for smaller training sets, while for the largest training sets temporal correlation was similar with no selection or causal selection.

The comparison of RMSE between predicted and observed rates under the different models and selection strategies is shown in Figures S3 and S4. Again causal selection appeared to improve predictions compared to no selection in the GP model. Using the linear model, no selection resulted in better performance for larger training set sizes and worse for smaller training sets. There are fewer clear trends when comparing causal and spike-and-slab selection.

\section{Discussion}
A notable feature of our results is the stability of the causal feature selection procedure across the different iterations when compared to spike-and-slab selection. Stability in feature selection is desirable for a number of reasons. Instability in a selection method, defined loosely as large changes in the selected feature sets arising from small perturbations of the data, undermines the idea that the selected feature set represents a meaningful set of driving factors. A consistently selected feature set, on the other hand, is more likely to reflect true associations between explanatory and response variables and therefore result in better out-of-sample predictions. Many common feature selection methods are known to be unstable when applied to linear or logistic regression~\citep{sauerbrei2015stability, kalousis2005stability, dunne2002solutions}. One reason for this is that given a large number of potential features (and therefore a large number of candidate models), using goodness-of-fit measures it is difficult to distinguish combinations of features that happen to fit the observed data well from truly informative feature sets. 

Bootstrapping samples has been proposed as a solution to this problem, for example the `little bootstrap' method proposed by~\citet{breiman1992little} to estimate predictive error, which is then used to compare models. \citet{austin2004bootstrap} proposed a bootstrapping selection method where backwards elimination selection was applied to bootstrapped samples of the data and, as in our method, the features were ranked by the number of times they were selected, with features that were selected in at least a fixed proportion included in the final feature set. Bootstrap-based adjustments are also appealing with regard to their potential improvements to the coverage of the Kullback-Leibler divergence minimising parameter set by the resulting Bayesian credible intervals in certain settings \citep{lyddon,huggins}.  However, in the regression context these methods do not address the consequences of a possible misspecification of the relationship between independent and dependent variables (such as resulting from incorrectly assuming a linear effect of each feature) in regard to the extrapolation to unobserved locations. If the model is well-specified, informative features should perform well (in terms of goodness-of-fit or predictive error) in many of the bootstrapped samples but in the presence of misspecification this is not necessarily the case. The stability of the causal feature selection procedure is likely to derive from both the use of bootstrapped samples and the non-parametric nature of the causal discovery algorithm. Spike-and-slab selection, on the other hand, showed relatively little consistency between iterations, suggesting that features were often selected due to chance associations in the training data that agreed with the model structure. While the more flexible GP model should suffer less from misspecification than the linear model, the assumed smoothness of these effects and lack of interactions between variables may still have prevented important variables being identified using spike-and-slab selection.

The ranking of variables by causal selection is also plausible---elevation is well-known to have a strong connection to transmission in Madagascar and aridity and PET are likely to influence the presence of mosquito breeding sites. Both \textit{P. falciparum} and \textit{P. vivax} are endemic in the country and therefore it is reasonable that the temperature suitability of both these species is an important factor. It is less clear why night time temperature would be a more causal factor than temperature during the day, although this could relate to night-biting behaviour of the mosquito vector. We would caution against interpreting the reasonableness of these results as validation of the causal selection procedure. However, the results could direct further investigation, for example looking at whether nighttime temperature is a more important driver of transmission in Madagascar than daytime temperature.

While the causal selection algorithm produced consistent rankings of dynamic variables, the relative importance of the different time lags for each variable was less clear (for example whether nighttime temperature at a 0, 1, 2, or 3 month time lag was most likely to be a direct cause). This is perhaps unsurprising, as there is likely to be variation in the time taken for a change in dynamic variable to affect the observed case count (due to variation in, for example, the duration of the infection process, time taken to seek treatment and reporting behaviour). However, it is also possible that the causal selection process was unable to infer meaningful differences between the same variable at different time lags and selected different time lags in different repeats based on noise. This may be a consequence of the stability of the algorithm, as we would expect a trade off between the stability and specificity of any selection procedure. 

As already discussed, spike-and-slab selection was much less stable than causal selection when applied to either the linear or GP model. There was some consistency in that static features were often found to be important, but otherwise there was little agreement between rankings in different iterations or when comparing the linear or GP models. When comparing the predictive ability of both models when using causal selection and spike-and-slab selection, using causal selection improved results in almost every context and otherwise produced comparable performance to using spike-and-slab selection. This is fairly remarkable, as the causal selection procedure was completely model-free whereas the feature sets generated by spike-and-slab selection were based on the model used. The improvement was clearest when using the GP model, in which case causal selection improved performance by every metric. This may be because the nonparametric causal selection procedure identified important features based on non-linear dependencies which were captured better by the GP regression model. 

When comparing causal selection to no feature selection, the linear model generally performed better with no selection, both in terms of overall correlation and temporal correlation. This suggests that there were features that were informative within the linear model framework that were not present in the causal feature sets. The exception to this was when using the smallest training sets, where model performance was better when using causal selection. In these low data settings, it is likely there is not enough information to learn effectively and avoid overfitting when using all features. In contrast, when using the GP model performance was better when using causal selection. Again, this may be because the causal selection procedure selected features based on non-linear relationships. Furthermore, overfitting may have been more of a problem when using all possible features in this more complex model. These results suggest that causal selection is likely to be beneficial in most settings and it is least effective in the high data, low complexity situations in which feature selection may not be needed at all. There are a number of other benefits to using smaller feature sets, including reduced computational requirements and improved interpretability. Collecting covariate information at new locations and updating and maintaining covariate data also comes at a cost, so building reliable models with small feature sets is beneficial. This is particularly true for models of disease risk which are likely to be rerun regularly with updated data. All together, this presents a strong case for using causal feature selection over not performing feature selection in many situations.

There are also a number of limitations to the proposed causal selection procedure. While this method scales well with the number of data points (due to the use of conditional independence tests that are of linear complexity), the PC algorithm itself is likely to scale poorly with large numbers of potential features.  More efficient causal discovery algorithms exist (see e.g.~\citealt{maathuis2009estimating, nandy2018high}) but require stronger assumptions about the data generating process. The PC algorithm may also be inappropriate due to the assumption that there are no unobserved confounders. In this case, however, our approach can be modified to use a discovery algorithm that is designed to be robust in the presence of some unobserved confounders, such as the FCI algorithm~\citep{spirtes2000causation}. We also note that unobserved confounders are more likely to led to edges being incorrectly retained, rather than incorrectly removed, in the PC algorithm and therefore the resulting parent set of the variable of interest is more likely to be a superset of the true set of parents than it is to be missing true parent variables. In many applications, including these extra variables in the feature set may not be a significant problem. This is in contrast to the problem of estimating causal effects, where the presence of unobserved confounders is more problematic. Finally, it is possible that the causal structure that the PC algorithm aims to infer is different in different parts of Madagascar due to the diverse ecological landscape. In this case, it may be more appropriate to run the selection procedure separately in different areas and aggregate the resulting feature sets somehow or to simply model different areas independently. Alternatively, if the functional relationships between the covariates response vary over space then this could be incorporated into the regression model structure. For example, spatially-varying coefficients could be used. For example, spatially-varying coefficients could be included. Such a model could be used with the causal selection procedure outlined here.

\section{Conclusion}
We have proposed a causal discovery procedure that can be applied to data with variables that have spatiotemporal structure and may or may not be aggregated over space and which places no parametric assumptions on the nature of these variables. Data with these properties are common in spatial epidemiology and other applied statistics problems. Our extension to kernel-based methods to accommodate aggregate data could be applied with any kernel-based independence or conditional independence test. We used this causal discovery algorithm to perform feature selection but causal discovery could be used as an exploratory tool or as an intermediate step to identifying causal effects. Kernel-based independence tests also have many other applications, such as independent component analysis, which could be applied to aggregated data using our proposed method.

Our results suggest that causal feature selection is a promising selection tool in disease mapping contexts. This selection method was highly stable across different training sets. Despite being completely model-free, causal selection in general led to better predictions than using spike-and-slab regression, a classical selection method. Causal selection also led to improved predictions over performing no feature selection except in high data settings with the less complex model. Causal selection was particularly advantageous for the more complex GP model and when training data sets were small.

\section*{Acknowledgements}
Arsene Ratsimbasoa and Thierry Franchard of the National Malaria Control Program of Madagascar are thanked for sharing routine malaria case data with the Malaria Atlas Project. Fanjasoa Rakotomanana and her team at the Institut Pasteur de Madagascar are also thanked for sharing the health facility geolocation data with us. Suzanne Keddie and Emma Collins are thanked for processing the data.

The first author was supported in this work through an Engineering and Physical Sciences Research Council (EPSRC) (https://epsrc.ukri.org/) Systems Biology studentship award (EP/G03706X/1). Work by the Malaria Atlas Project on methods development for Malaria Eradication Metrics including this work is supported by a grant from the Bill and Melinda Gates Foundation (OPP1197730).

\appendix

\FloatBarrier
\bibliographystyle{rss}

\end{document}